\documentclass[twocolumn]{aastex63}

\usepackage{amsmath}
\usepackage{amssymb}
\usepackage{euscript}
\usepackage{bbold}
\usepackage{paralist}
\usepackage{color}
\usepackage{amsmath}
\usepackage{lineno}
\usepackage{lineno}
\usepackage{breqn}

\usepackage{comment}

\submitjournal{PASP}

\shorttitle{X-Sifter}
\shortauthors{Soumagnac et al.}

\begin{document}


\title{{\tt X-Sifter}: detecting transients in X-ray data using the optimal Poisson matched filter}


\correspondingauthor{Maayane T. Soumagnac}
\email{mtsoumagnac@lbl.gov}
\author[0000-0001-6753-1488]{Maayane T. Soumagnac}
\affiliation{Department of Physics, Bar-Ilan University, Ramat-Gan 52900, Israel}
\affiliation{Lawrence Berkeley National Laboratory, 1 Cyclotron Road, Berkeley, CA 94720, USA}
\author{Eran O. Ofek}
\affiliation{Benoziyo Center for Astrophysics, Weizmann Institute of Science, 76100 Rehovot, Israel}
\author{Shachar Israeli}
\affiliation{Department of Physics, Bar-Ilan University, Ramat-Gan 52900, Israel}
\author{Guy Nir}
\affiliation{Lawrence Berkeley National Laboratory, 1 Cyclotron Road, Berkeley, CA 94720, USA}
\affiliation{Benoziyo Center for Astrophysics, Weizmann Institute of Science, 76100 Rehovot, Israel}
\author{Imri Dickstein}
\affiliation{Department of Physics, Bar-Ilan University, Ramat-Gan 52900, Israel}


\begin{abstract}

We present {\tt X-sifter}, a software package designed for near-optimal detection of sources in X-ray images and other forms of photon images in the Poisson-noise regime. The code is based on the Poisson-noise-matched filter (Ofek \& Zackay), which provides an efficient method for calculating the delta log-likelihood function for source detection.
The software accounts for several complexities inherent in real data, including variations in both the instrumental Point Spread Function (PSF) and background across the detector and as a function of energy. We validate the pipeline using real data with simulated source injections, as well as actual {\it Chandra} images. A comparison between the sources detected by our pipeline and those in the Chandra Source Catalog (CSC) suggests an approximate $\cong30\%$ increase in the number of detected (real) sources. Near the detection limit, the reported $S/N$ of our pipeline is approximately $1.3\times$ higher than that of the CSC. This corresponds to a factor of $1.8$ increase in survey speed.

\end{abstract}


\section{Introduction}

Astronomical image processing typically begins with a source detection step. In the case of independent and identically distributed (i.i.d.) Gaussian noise, this problem has an established optimal solution known as the {\it matched filter} method. This approach is widely used in both science and engineering (e.g., \citealt{Stetson1987_DAOPHOT,Bertin+1996_SExtractor,Ofek+2023PASP_LAST_PipeplineI}; see review in \citealt{Zackay+2017_CoadditionI}).
The matched filter can be derived using the lemma of Neymann-Pearson (\citealt{Neyman+Pearson1933_HypothesisTesing}), and it is a simple hypothesis testing between the null hypothesis $\mathcal{H}_0$ (that there is no source in a specific position) and the alternative hypothesis $\mathcal{H}_1$ (that there is a source in that position). In the Gaussian-noise case, the solution involves cross-correlating (filtering) the observed signal with the Point Spread Function (PSF) of the image, followed by the identification of local maxima in the filtered image. Since cross-correlation can be performed efficiently using the Fast Fourier Transform (FFT), this method is not only optimal but also efficient.

However, in the low number counts regime, when the photon statistics follow Poisson statistics, the Gaussian matched filter becomes sub-optimal. 
This regime is relevant to several astronomical applications, such as the detection of sources in low-count imaging (X-ray, UV, $\gamma$-ray, and neutrino detectors) and counts-in-cell. 

Some of the most widely used source detection techniques in X-ray astronomy are {\tt wavedetect} \citep{Freeman2002} and the sliding-cell method \citep{Harnden1984} (for a comprehensive overview, see \citealt{Masias2012}). Both methods fall under the category of filtering techniques, where a filter (or Kernel) is cross-correlated with the image: {\tt wavdetect} cross-correlates the image with a wavelet function, while the sliding-cell method employs a top-hat function for this purpose. However, these filters are suboptimal in the Poisson-noise regime, meaning that cross-correlating with them results in detections with lower signal-to-noise ratios ($S/N$) compared to the theoretical maximum.

One approach to addressing the source detection problem in the low-count regime involves calculating the likelihood function. Methods based on this approach include: (1) the Cash statistic \citep{Cash1979}, implemented in tools such as the {\tt X-Spec} pipeline\footnote{\url{https://heasarc.gsfc.nasa.gov/xanadu/xspec/}} for X-ray spectral fitting, and (2) the likelihood analysis widely employed in many Fermi data analyses \citep{Mattox1996, Macias2018, Fermi2023}. However, likelihood-based methods can be computationally intensive.

To address these challenges, \cite{Stewart2006} proposed a semi-heuristic adaptation of the traditional matched filter for the non-Gaussian case. Additionally, \cite{Vio2018_MatchedFilter_LooksElseWhere} suggested employing the Saddle-Point approximation to simplify the computation of the probability distribution for detecting a signal under the null hypothesis ($\mathcal{H}_0$).

\cite{Ofek+2018_PoissonNoiseMF} adopted a different approach, demonstrating that the problem of source detection in the low number counts regime can be formulated analogously to the matched filter approach, i.e. as a filtering problem. In their method, detection statistics (specifically, the delta-log-likelihood function) are derived by cross-correlating the signal with a slightly modified version of the kernel used in the Gaussian case. The advantage of this method is that the score image can be computed efficiently using FFT,eliminating the need to calculate the likelihood function explicitly at each pixel. 

In this paper, we present {\tt X-Sifter}, an
implementation of the Poisson Matched Filter (PMF) formalism proposed by \cite{Ofek2018}. We adapted this formalism into an algorithm designed to handle real data, accounting for various complications such as background and PSF variations across the detector.
\S\ref{sec:poisson} provides an overview of the key aspects of the PMF formalism, including corrections to the thresholding equations and strategies adopted to address various real-data challenges inherent to the method. The main steps of the {\tt X-Sifter} algorithm are detailed in \S\ref{sec:algo}. In \S\ref{sec:results}, we validate the pipeline with simulated sources and compare its sensitivity to that of sources detected in the Chandra Source Catalogue.

\section{Matched filtering in the low-number count Poisson noise regime}\label{sec:poisson}

\subsection{The Poisson Matched filter formalism}

%

The {\tt X-Sifter} algorithm is based on the optimal matched filter in the low-number count Poisson noise regime derived by \cite{Ofek2018}. Here, we summarize the main aspects of the method and refer to \cite{Ofek2018} for more details. We also generalize some aspects of the formalism. 

Declaring a detection in the regime of low-number count Poisson noise involves calculating the statistic $S$, the log-likelihood difference between the null hypothesis $\mathcal{H}_0$ that there is no source in a given pixel position (in this case $M$, the measured 2-D image at location $q$ is best described by pure background noise $B$, assumed to be constant across all bins: $M(q) \sim \text{Poisson}(B)$), and the alternative hypothesis $\mathcal{H}_1$ that flux from a source contributes to the measured image in that location (in this case $M(q) \sim \text{Poisson}(B + F_s P(q-q_0))$, where $P$ is the kernel normalized to unity, $q_0$ is the location of the source, and $F_s$ is the source flux). \cite{Ofek+2018_PoissonNoiseMF} derived the expression for $S$ and wrote it using the cross-correlation operator:
%
%
\begin{equation}\label{eq:cross}
S=M\otimes\overleftarrow{\mathcal{K}_{PMF}}\;.
\end{equation}
Here $\otimes{}$ denotes convolution, $\xleftarrow{}$ denotes coordinates inversion ($q\xrightarrow{} -q)$, and $\mathcal{K}_{PMF}$ is the Poisson noise optimal filter given by:
\begin{equation}\label{eq:PMF}
    \mathcal{K}_{PMF}(F_s)=\ln{(1+\frac{F_{s}}{B}P)}\;.
\end{equation}

\noindent Equation~\ref{eq:cross} can also be written in Fourier space:
\begin{equation}
    S=\widehat{M}\overline{\widehat{\mathcal{K}_{PMF}}}\;,
\end{equation}
where the $\widehat{\phantom{X}}$ symbol denotes the Fourier transform and the bar symbol denotes complex conjugation.

The lemma of Neymann-Pearson (\citealt{Neyman+Pearson1933_HypothesisTesing}) states that the likelihood-ratio test which rejects the null hypothesis $\mathcal{H}_0$ in favor of $\mathcal{H}_1$ when $S\geq S_{th}$ is the most powerful test, at a constant false-alarm probability level 
$\gamma=\mathcal{P}(\Lambda\geq S_{th}\mid H_0)$ and a threshold $S_{th}$, defined by 
\begin{equation}\label{eq:Sth}
    \gamma=\int_{S_{th}}^{\infty} \mathcal{P}(S\mid \mathcal{H}_0)dS\,.
\end{equation}

In the well-known Gaussian-noise case, the filter (or Kernel) $\mathcal{K}$ is simply the PSF, and the flux $F_s$ of the source is absorbed into the threshold $S_{th}$, whereas in the Poisson case, the threshold depends on $F_s$ in a non-trivial way, as detailed in the next section.

\subsection{Complications}

Several complications, discussed in Sections~\ref{sec:thresh}, \ref{sec:ps}, \ref{sec:gamma-S} are directly related to the Poisson matched filter formalism. Additional challenges, addressed in Sections~\ref{sec:dectector_sector} and~\ref{sec:channeling}, arise from instrumental and real-data complexities.

\subsubsection{Calculation of the threshold $S_{th}$ and thresholding relation}
\label{sec:thresh}

Unlike in the Gaussian noise case, in the Poisson noise case, the Kernel $\mathcal{K}_{PMF}$ (see Equation~\ref{eq:PMF}), and subsequently the threshold $S_{th}$ used to determine whether a source has been detected, depends on $F_s$, the flux of the source, which is not known a priori.
In statistical terms, $\mathcal{H}_1$ depends on the free parameter $F_s$, meaning it is not a simple hypothesis as required by the Neyman-Pearson lemma. 

\begin{figure}
\includegraphics[width=8cm]{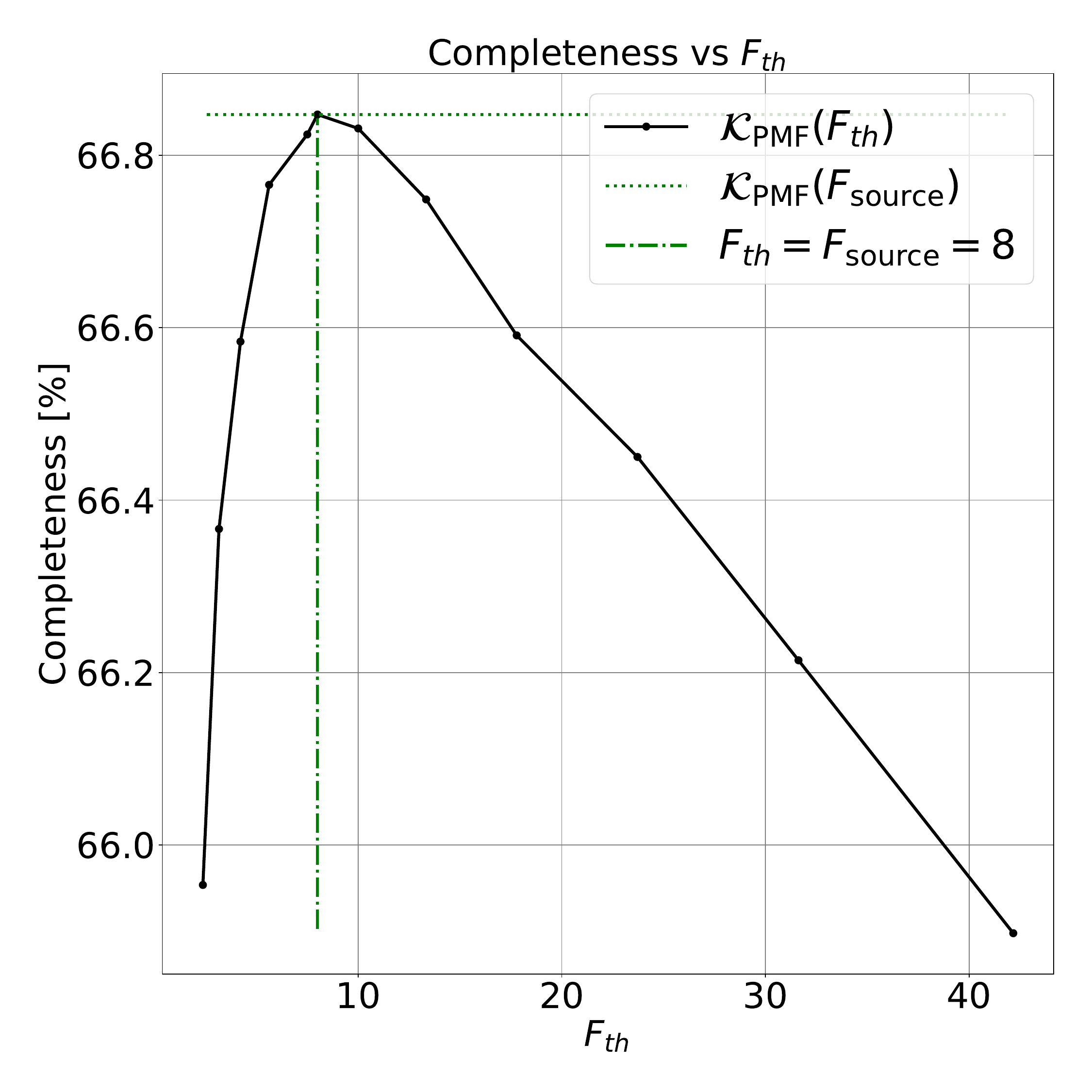}
\caption{{\bf The effect of choosing a constant flux $F_{th}$ in the PMF, studied in 1-D}. A source with a flux $F_{s}$ is simulated and filtered with the approximated PMF Kernel of Equation~\ref{eq:kpmf_fth}, with various values of $F_{th}$. The y-axis shows the completeness of the detection for each choice of $F_{th}$. Although the completeness is maximal when $F_{th}=F_{s}$, it is affected by less than $1\%$ when $F_{th}$ varies by an order of magnitude. See also \cite{Ofek+2018_PoissonNoiseMF}. The parameters used to produce this figure are the following: $B = 0.0125\,\rm photons/\rm bin$, $\sigma_{th} = 5$, $F_{\text{source}} = 8 \,\rm photons$. P is a FRED template spanning 75 bins. All $\mathcal{P}(S\mid H_0)$ and $\mathcal{P}(S\mid H_1)$ distributions used in calculating the completeness values are simulated with $10^{10}$ simulations. Note that the completeness is higher than $50\%$ at peak because the chosen $F_s=8\,\rm photons$ is higher than the detection limit, for this value of the background.}
\label{fig:Fth}
\end{figure}

In Figure~\ref{fig:Fth}, A source with a flux $F_{s}$ is simulated and filtered with the PMF Kernel $\mathcal{K}_{PMF}(F)$, for a range of values of $F$. The y-axis shows the completeness\footnote{the completeness is the probability \begin{equation}
     \int_{S_{th}}^{\infty} \mathcal{P}(S \mid H_1) dS\,.
\label{eq:completeness}
\end{equation} for real sources to be detected above $S_{th}$.} of the detection for each choice of $F$. While completeness is maximized when $F=F_s$, the Figure shows it is only slightly affected by variations in the value of $F$ (less than $1\%$ when $F$ varies by an order of magnitude). Therefore, choosing a fixed value $F_{th}$ and filtering the entire image with
\begin{equation}\label{eq:kpmf_fth}
\mathcal{K}_{PMF}(F_{th})=\ln(1+\frac{F_{th}}{B}P)\,,
\end{equation}
results in only minor losses in completeness, as long as this fixed value $F_{th}$ is chosen to be in the general ``ballpark'' of the source flux. In other words, the statistic from Equation~\ref{eq:cross} can be approximated by:
\begin{equation}\label{eq:approx}
S\cong M  \otimes   \overleftarrow{\mathcal{K}_{PMF}(F_{th})}\,
\end{equation}
and gives the best performance when $F_{th}=F_{s}$

%

The {\it thresholding relation} is the requirement that the method be optimized for the faintest sources we are interested in, i.e. such that $F_{th} = F_s$ at the detection limit. By definition, at the detection limit, a source with flux $F_s$ has approximately a 50\% probability of exceeding the threshold $S_{th}$ and being detected. This definition implies that the corresponding distribution $\mathcal{P}(S \mid \mathcal{H}_1)(F_s)$ under the alternative hypothesis $\mathcal{H}1$ should peak near $S_{th}$:

%


\begin{equation}\label{eq:condition}
S_{th}=\mathbb{E}\left(S_{\mathcal{H}_1}(F_s)\right),
\end{equation}
where $\mathbb{E}$ is the expectancy value. Since
 \begin{align}\label{eq:average}
\mathbb{E}(S_{\mathcal{H}_1}(F_s)) =\mathbb{E}\left(M_{\mathcal{H}_1} (q) \otimes \overleftarrow{\mathcal{K}_{PMF}(q)}\right) \\
    = \mathbb{E}\left(\text{Poisson}\left(B(q) + F_sP(q)\right) \otimes \overleftarrow{\mathcal{K}_{PMF}(q)}\right)\;,
\end{align}

\begin{figure}
\includegraphics[width=9cm]{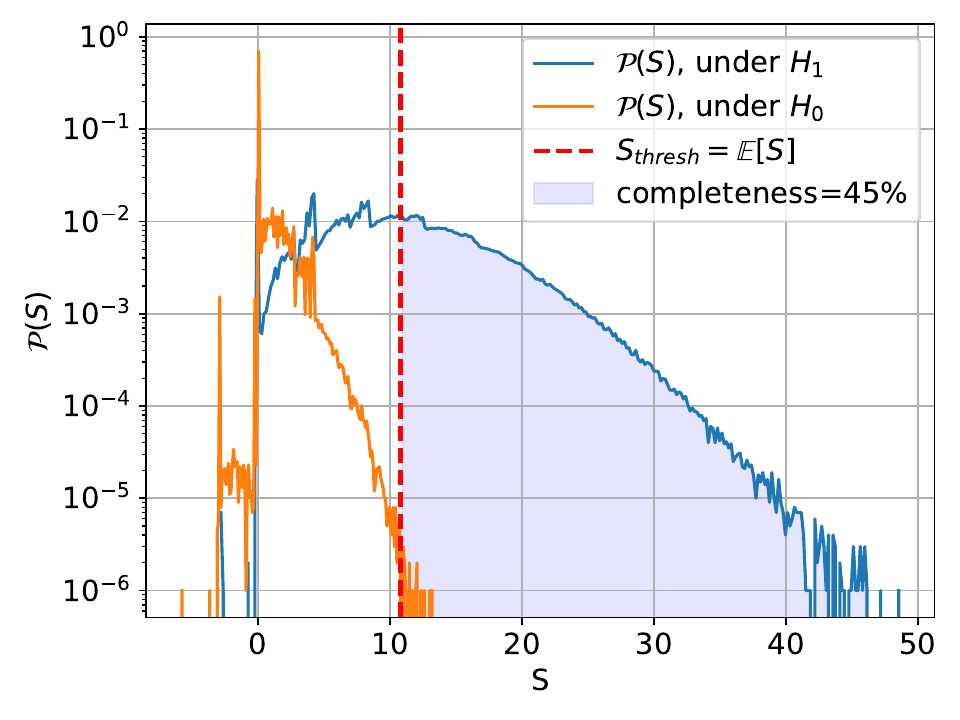}
\caption{{\bf The thresholding relation is equivalent to requiring $F_{th}=F_{s}$ at the detection limit.} For a given $P$ and $B$, we calculated $\mathcal{P}(S\mid H_0)$, $S_{th}$ and then derived $F_{th}$ from Equation~\ref{eq:final link}. We then simulated $10,000$ images, each containing a single source with flux $F_{s}=F_{th}$. We filtered these images using the optimal kernel $\mathcal{K}_{PMF}(F_{th})=\mathcal{K}_{PMF}(F_s)$. The resulting distribution $\mathcal{P}(S\mid \mathcal{H}_1)$ is shown in blue, while the orange line represents $\mathcal{P}(S\mid H_0)$. $\mathcal{P}(S\mid \mathcal{H}_1)$ peaks at $S=S_{th}$ meaning that we are at the detection limit: about half the cases with expectancy value $F_s=F_{\rm th}$ fall above $S_{th}$.}
\label{fig:P_H0_H1}
\end{figure}
Equation~\ref{eq:condition} translates into
\begin{equation}
    S_{th} = B(q)\otimes\overleftarrow{\mathcal{K}_{PMF}(q)}+F_s P(q) \otimes \overleftarrow{\mathcal{K}_{PMF}(q)} \;,
\end{equation}
i.e.
\begin{equation}\label{eq:link}
    S_{th} = B\sum_q{\ln{\left[1+\frac{F_s}{B}P(q)\right]}} + F_s \cdot \mathcal{C}(F_s)\;,
\end{equation}
where
\begin{equation}\label{eq:SF}
    \mathcal{C}(F)=\sum_q{P(q)\ln{\left[1+\frac{F}{B}P(q)\right]}}\,.
\end{equation}

Given $S_{\rm th}$, $B$, and $P$, requiring that $F_{th} = F_s$ at the detection limit is equivalent to determining a self-consistent fixed value of $F_{th}$ that satisfies Equation~\ref{eq:link}. This condition holds when:

\begin{equation}\label{eq:final link}
S_{th} = B \sum_q \ln{\left[1 + \frac{F_{th}}{B} P(q)\right]} + F_{th} \cdot \mathcal{C}(F_{th})\;.
\end{equation}

This equation linking $S_{th}$ and $F_{th}$, which we refer to as the {\it thresholding relation}, is a generalization of the relation proposed in \cite{Ofek2018} and can be solved numerically. 
The equivalence between Equation~\ref{eq:final link} and the condition expressed in Equation~\ref{eq:condition} is illustrated in Figure~\ref{fig:P_H0_H1}, in which we simulated $10,000$ images, each containing a single source with flux $F_{s}=F_{th}$, where $F_{th}$ was derived from Equation~\ref{eq:final link}. Upon filtering these images with $\mathcal{K}_{PMF}(F_{th})$,
we show that $\mathcal{P}(S\mid \mathcal{H}_1)$ peaks at $S=S_{th}$ (Equation~\ref{eq:condition}), meaning that about half the cases with expectancy value $F_s=F_{\rm th}$ fall above the detection limit. 

\subsubsection{Computation of $\mathcal{P}(S\mid H_0)$}\label{sec:ps}

Declaring a detection assumes that we can compare the values of the filtered image $S$ to the threshold value $S_{th}$ defined in Equation~\ref{eq:Sth}. This involves being able to compute $\mathcal{P}(S\mid H_0)$, which is a complicated distribution with dips and peaks (shown e.g. in the right panels of Figure~\ref{fig:PSF1PSF2}).

Although attempts to compute $\mathcal{P}(S\mid H_0)$ analytically have been made \citep{SPA-PMF-Vio-Andreani}, the approach advocated by \cite{Ofek2018} and implemented here is to estimate $\mathcal{P}(S\mid \mathcal{H}_0)$ using numerical simulations. Each simulated $S_{\rm sim}$ value is obtained as the central pixel of the 
cross-correlation of a simulated background image, $B_{\rm sim}$, generated using a Poisson distribution with an expected value $B$, and the PMF kernel:

\begin{equation}\label{eq:Ssim}
    S_{\rm sim}=\sum_{x}\sum_{y}B_{\rm sim}\overleftarrow{{\mathcal{K}_{PMF}}}\;.
\end{equation}

(Note that there is no need to do full cross-correlation in order to obtain $S_{sim}$, since we are interested only in the central pixel).
To improve runtime, $\mathcal{P}(S\mid H_0)$ can be calculated once, for a library of PSF stamps and a grid of background values, as will be detailed in \S\ref{sec:modelingS}.

\subsubsection{The $\gamma-S_{thresh}$ relation}\label{sec:gamma-S}

For values of the false-alarm probability $\gamma$ corresponding to $S/N\gtrsim5\sigma$, the number of simulations required to properly sample $\mathcal{P}(S\mid H_0)$ becomes prohibitive. One solution, proposed by \cite{Ofek2018}, is to fit the $\gamma-S_{th}$ relation in regions where it is well-sampled, based on the assumption that it follows an exponential law ($\log(\gamma)\propto S$), and then extrapolate this dependency for small values of $\gamma$. In \cite{Ofek2018}, the authors suggested that the $\gamma-S_{th}$ relation can be universally approximated by an exponential law with a slope of approximately $-1/2$. However, our simulations using real Chandra PSF stamps indicate that the exact shape of the  $\gamma-S_{thresh}$ relation depends on both the background level and the PSF and that (1) it is not always well approximated by an exponent (2) even in the regime of background noise where $\log(\gamma)\propto S$ is a good approximation, we observe a range of slope ranging from $\sim -1/2$ and $\sim -3/2$ for the Chandra PSF library and the range of background values we modeled.

\begin{figure}
    \centering
    \includegraphics[width=9cm]{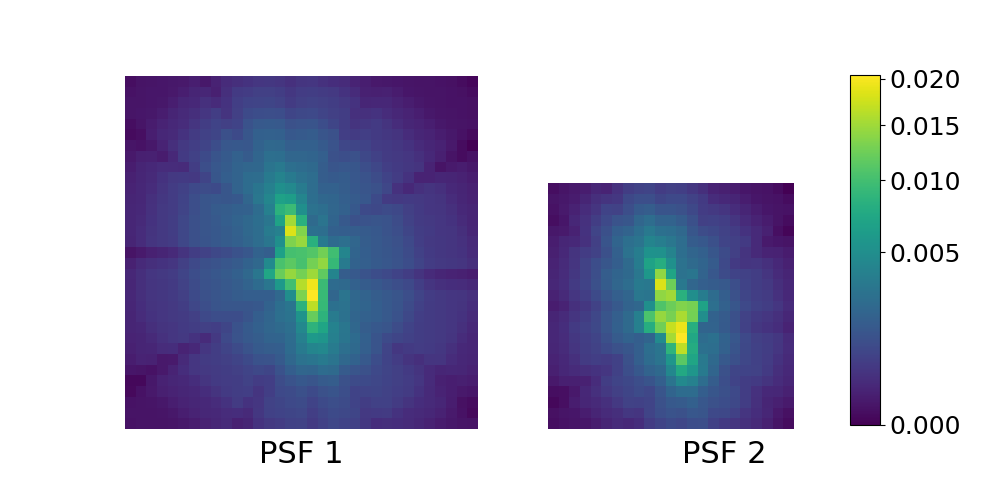}
    \includegraphics[width=8cm]{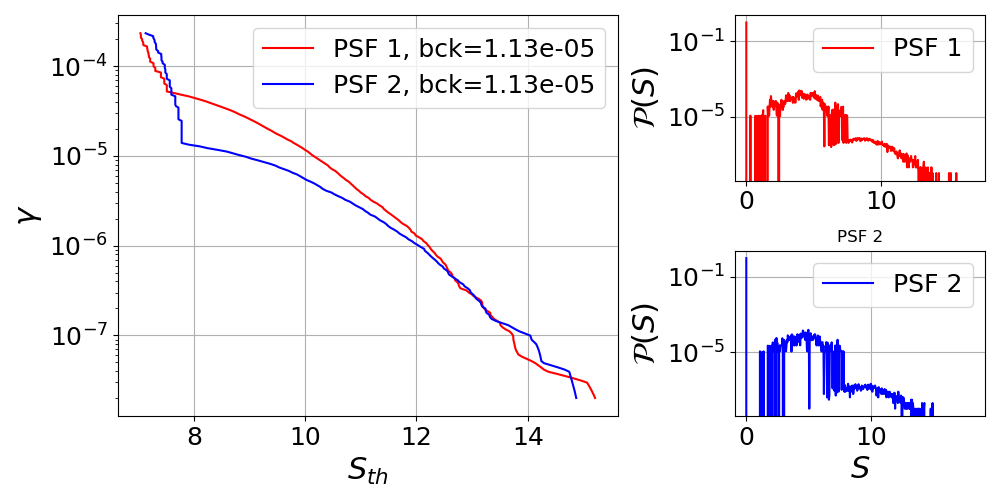}
    \includegraphics[width=8cm]{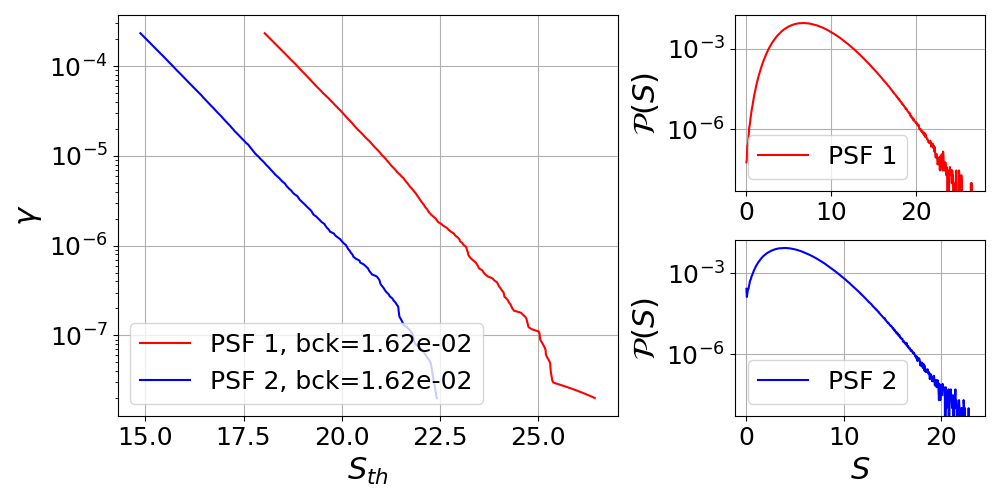}
    \caption{{\bf The effect of various background values and various PSF on $\mathcal{P}(S\mid H_0)$ and the $\gamma-S$ relation}. The top panel shows two PSF stamps, labeled PSF1 and PSF2, modeled with {\tt MARX} at two different detector location of CCD0 of the ACIS Chandra camera. The middle panels show the $\gamma)-S$ relation (left) and the $\mathcal{P}(S\mid H_0)$ distribution computed with PSF1 (in red) and with PSF2 (in blue), for a background level  $bck=1.13\times 10^{-5}$\,counts. The lower panels show the same for a background level $bck=1.62\times 10^{-2}$\,counts. At low background values, $\mathcal{P}(S\mid H_0)$ is full of dips and peaks, resulting in a complicated $\gamma)-S$ relation showing steps. As the background level grows, $\mathcal{P}(S\mid H_0)$ becomes smoother and the $\log(\gamma)$ becomes linear in $S$.}
    \label{fig:PSF1PSF2}
\end{figure}
\begin{figure}
\includegraphics[width=8cm]{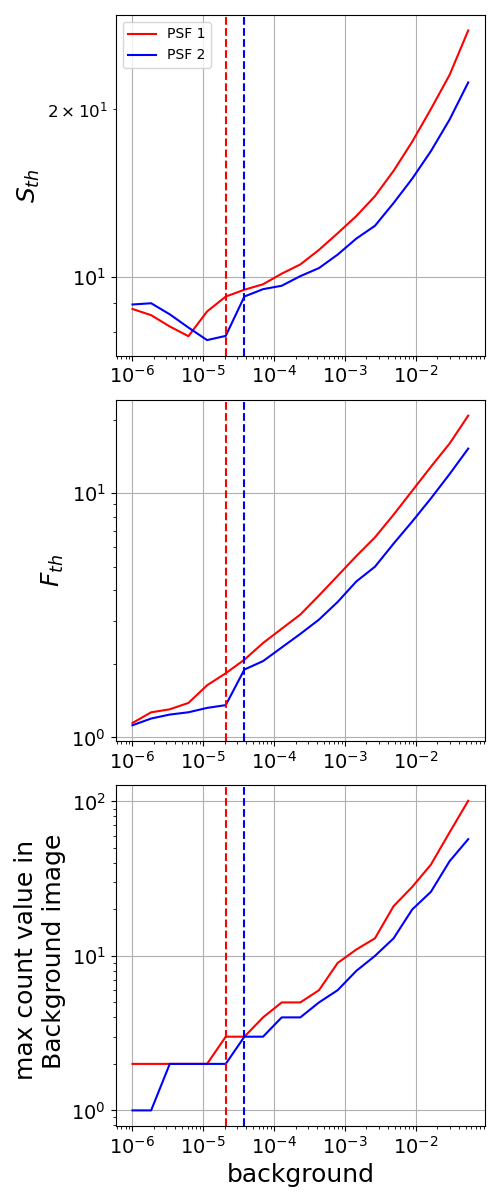} 
\caption{For each background values, images $B_{sim}$ where simulated (see ~\S\ref{sec:ps}) and used to compute $\mathcal{P}(S\mid H_0)$ and derive $S{th}$ (see Equation~\ref{eq:Ssim}). The bottom panel shows the maximum photon count value recorded in the $B_{sim}$ images. All quantities were computed using the stamps PSF1 (red) and PSF2 (blue), also shown in the top panel of Figure~\ref{fig:PSF1PSF2}. For low background levels, when the pixel counts in $B_{sim}$ are typically $0$, $1$, or $2$ photons, the $F_{th}$ values plateau, and $S_{th}$ exhibits a local minimum. At higher background levels, as the range of accessible photon counts values in the $B_{sim}$ image broadens, both $F_{th}$ and $S_{th}$ increase monotonically with the background level. Dashed vertical lines mark the region where $B_{sim}(q) \leq 2$, highlighting the transition between the plateau (in $Fth$) or dip (in $S_{th}$) and the monotonic increase phase.}
    \label{fig:bck_Sth_Fth}
\end{figure}

In Figure~\ref{fig:PSF1PSF2}, we present $\mathcal{P}(S \mid H_0)$ and the $\gamma-S_{th}$ relation computed for two background levels and two PSF stamps, modeled for one of the Chandra CCDs. Figure~\ref{fig:bck_Sth_Fth} illustrates the evolution of $F_{th}$, $S_{th}$, and the maximum photon count in the simulated images $B_{sim}$ as functions of the background expectancy.

At low background levels (middle panels of Figure~\ref{fig:PSF1PSF2}, left-hand side of Figure~\ref{fig:bck_Sth_Fth}), the distribution $\mathcal{P}(S \mid H_0)$ shows pronounced peaks and dips, resulting in a non-smooth, step-like $\gamma-S_{th}$ relation. The low photon counts in $B_{sim}$ (bottom panel of Figure~\ref{fig:bck_Sth_Fth}) lead to an initial plateau in $F_{th}$ (middle panel of Figure~\ref{fig:bck_Sth_Fth}) and a local minimum in $S_{th}$ (top panel of Figure~\ref{fig:bck_Sth_Fth}).

At higher background levels (lower panels of Figure~\ref{fig:PSF1PSF2}, right-hand side of Figure~\ref{fig:bck_Sth_Fth}), the distribution $\mathcal{P}(S \mid H_0)$ becomes smoother and the $\log(\gamma)-S_{th}$ relation transitions to a linear form (i.e. $\log(\gamma) \propto S$)
and the evolution of $F_{th}$ and $S_{th}$ with the background becomes monotonic.

The complexity of the $\gamma-S$ relation adds difficulty to converting the filtered image (in units of $S$) into detection significance units (in units of $\gamma$ or $\sigma$). To address this, we precompute this relationship for a grid of $B$ values and a library of PSF stamps (\S\ref{sec:modelingS}).

\subsubsection{PSF and background variation on the detector}
\label{sec:dectector_sector}
In X-ray images, the PSF and background, two key ingredients of the PMF Kernel (see e.g. Equation~\ref{eq:PMF}) can vary substantially across the field of view. One way to address this challenge is to partition the images into regions where the PSF and background are approximately constant and then filter each sub-image separately. The size of these regions is determined by estimating the information loss due to the use of the wrong PSF/background. The PSF of X-ray telescopes often has complex shapes with significant substructures. Figure~\ref{fig:PSF1PSF2} shows Chandra PSF profiles modeled at two different locations on the same Chandra CCD. In \S\ref{sec:psf_all}, we describe how the PSF is calculated in practice, ensuring that the computational runtime remains reasonable.

\subsubsection{Energy dependence}\label{sec:channeling}
Often in X-ray images, both the PSF and the background are energy-dependent.
Ignoring this will result in a loss of sensitivity.
To solve this problem we divide the image into 
multiple ($n_{\rm c}$) energy channels. The filter $\mathcal{K}_{PMF}$ (Equation~\ref{eq:PMF}) is calculated separately in each energy-channel and each energy-channel image is filtered with its own filter. This approach is valid as long as the energy range in which the PSF and background change is larger than the uncertainty in the photons energy. Finally, as the energy channels provide independent information, the overall signal-to-noise can be easily derived, by combining the signal-to-noise in the $n_c$ separate channels through\footnote{See \citealt{Zackay+2017_CoadditionI} for justification.}:
\begin{equation}\label{eq:combine}
(S/N)_{tot}=\sqrt{\sum_{i}^{n_{\rm c}} (S/N)_{i}^{2}}
\end{equation}
In \S\ref{sec:overview}, we explain how the S/N is calculated in practice.

\subsection{Temporal subdivision of the data}\label{sec:temporal}

Transients with a brief duration that only rise above the detection threshold for a very short time, can easily become buried within the background noise that accumulates during extended exposures. This phenomenon explains why many transients go unnoticed in many large source catalogs, e.g., in the CSC. This fact is aggravated by the exposures being stacked before being processed, which enhances sensitivity to faint, quiescent sources, but further diminishes the ability to detect short-duration signals. A vital step in detecting X-ray transients is thus to subdivide all data into shorter exposures before applying the search algorithm. This important step – which was incorporated into the studies by \cite{Law2004}, \cite{Alp2020} and \cite{DeLucas2021} which specifically searched transient events, increases the number of observations to process. The algorithm must run multiple times -- once for each subdivision of an observation -- introducing a runtime challenge. Our algorithm, {\tt X-Sifter}, includes a step that can subdivide the entire data into sub-exposures, optimized for the detection of the transients with the desired duration. 

In this context, one question is: what sub-exposure time should we use?
There is no benefit in arbitrarily decreasing the sub-exposure time, because when the exposure is very short, the expected value of the background becomes so small that any detected photon is likely to be associated with a source.
The choice of sub-exposure time depends on the number of false alarms we are willing to allow and the minimal number of photons required from a source.

For example, if we require that a source must have at least five photons and that the false alarm probability is below $10^{-13}$, Poisson statistics imply that the expected value of the background should be approximately $\approx 0.01$ in the area of the PSF.
Given the typical Chandra ACIS background rate of $8\times10^{-6}$ photons per second for an area of about 30 pixels, it suffices to divide the images into sub-images with an exposure time of about $1000\,\mathrm{s}$.

\section{The {\tt X-Sifter} algorithm for source detection in images}
\label{sec:algo}

\subsection{Algorithm overview}\label{sec:overview}
\begin{figure*}
\centering
\includegraphics[width=16cm]{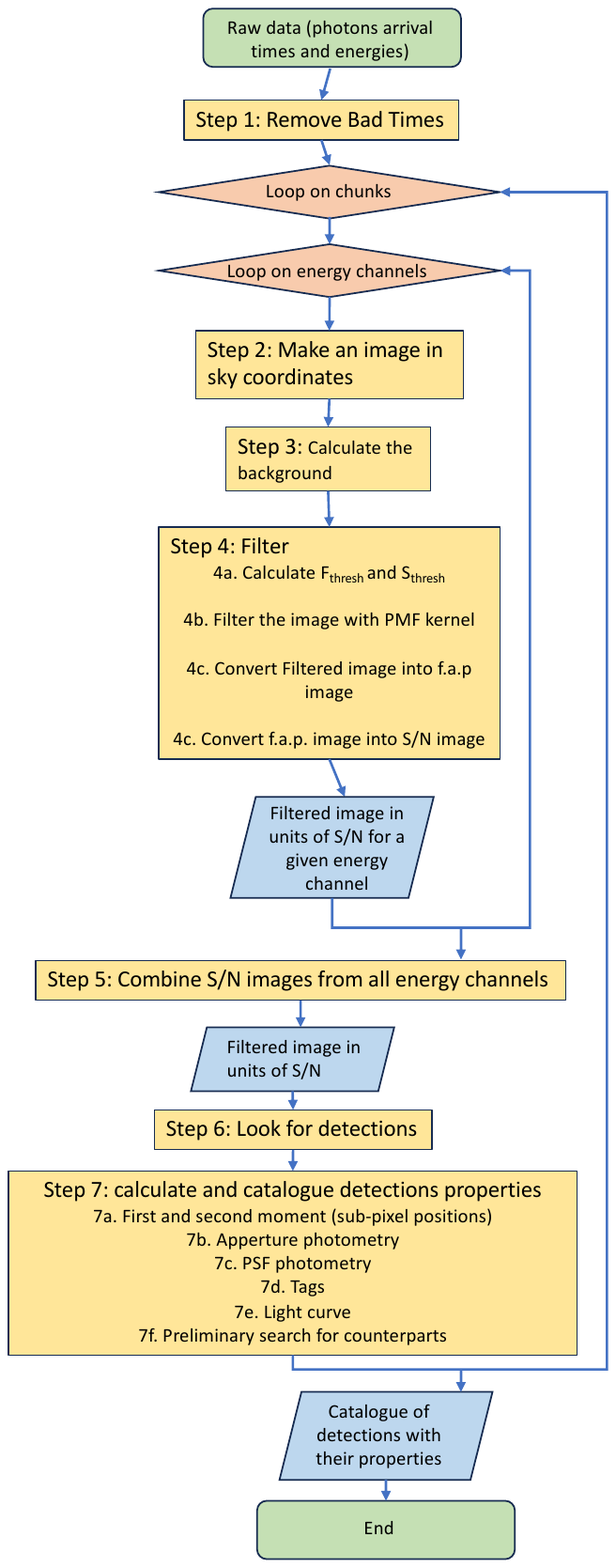}
\caption{{\bf Description of all the steps of the {\tt X-Sifter} algorithm}. Yellow boxes designate actions, blue diamonds designate data outputs/inputs, red diamonds signal loops and the green boxes are for the input and output of the pipeline.
}

\label{fig:flowchart}
\end{figure*}

The {\tt X-Sifter} pipeline has been written for Chandra ACIS imaging data, but it can be adapted to run on other X-ray archives, such as {\it XMM-Newton}.
The pipeline basic routine runs on a single observation, or {\tt OBSID}. 
For each CCD, the following steps are performed, and summarized in Figure~\ref{fig:flowchart}. 

\begin{enumerate}
    \item We exclude from the data any ``bad times'' during which a solar flare illuminated the entire CCD. We use a routine inspired by the MATLAB routine {\tt find_badtimes} by \cite{OfekMatlab2014}, which identifies and rejects time windows where the mean count rate exceeds the median count rate for the entire observation by a chosen number of standard deviations. 
    (By default, windows of time with $100$ events, on average, with background higher than $4$ standard deviation above the median count rate are rejected).
    
    \item We filter out all the photons below $0.2\,\rm keV$ and above $8\,\rm keV$. This step helps account for the Chandra degradation over the last decade. The selected band of energies is an adjustable parameter of the pipeline.
    \item We partition each CCD of the selected {\tt OBSID} into $16 \times 16$ squares, referred to as ``sectors'', to ensure that both the PSF and the background remain relatively constant across the area where the image is filtered (see \S\ref{sec:dectector_sector}). A ``buffer'' of $160$ pixels is added to each sector, overlapping with adjacent sectors, to accommodate filtering with potentially large PSF stamps, even along the sector borders.
    \item We partition the photons sample into $n_c$ energy channels, logarithmically spaced (in keV units): e.g. when $n_c=3$, the channels are: $[0.2,0.68399 ]$, $[0.68399,2.3392]$ and $[2.3392,8]$ (see \S\ref{sec:channeling}). We then loop through both energy channels and sectors, executing the following tasks:
    
    \begin{enumerate}
        \item We either model the PSF for the specific sector and energy channel (see section\,\ref{sec:psf}), or use a pre-modeled PSF stamp from a PSF library we computed (see section\,\ref{sec:library}) which is then rotated by the roll angle of the observation to match the PSF at the desired location.
         \item We make an image in {\tt sky}  coordinates by binning the data, into the $0.492"$ Chandra pixels. 
         \item We calculate an estimate of the background for this image. This is done using a routine inspired by \cite{OfekMatlab2014}. The image is divided into a grid of $n\times n$ square tiles (by default $n=4$). For each tile, the code calculates the mean photons count (as an initial background estimate) and the variance and then iteratively refines the background estimate by excluding high-intensity pixels (above certain thresholds like the 95th percentile or Poisson limits), recomputing the mean and variance for the remaining pixels until the background estimate stabilizes. The function then produces an array containing the background values for each tile and computes the median background value as the representative background for the entire sector.
         \item We solve Equation~\ref{eq:final link} to calculate $F_{th}$ for the Poisson matched-filter corresponding to PSF computed in (a), the background $B$ calculated in (b)and a chosen false-alarm probability $\gamma$. 
         \item We compute $\mathcal{P}(S\mid H_0)$ (see \S\ref{sec:ps}) and deduce $S_{th}$ by taking the percentile corresponding to the value of $\gamma$ chosen in (d).
         \item From $\mathcal{P}(S\mid H_0)$ conputed in (e), we derive the $\gamma - S$ relation, using $\gamma(S) = 1-{\tt cumsum}\left(\mathcal{P}(S\mid H_0)\right)$ where {\tt cumsum} is the cumulative sum. 
         \item We compute the Poisson matched filter kernel $\mathcal{K}_{PMF}$ using Equation~\ref{eq:kpmf_fth}.
         \item We filter the sky-coordinates image computed in (b) with the PMF Kernel calculated in (g) (Equation~\ref{eq:approx}), using FFT.
         \item We convert the filtered image (units of S) into an image in units of $\gamma$ using the relation derived in (f).
         \item We convert the image in units of $\gamma$ into an image in units of ``Gaussian'' S/N, using $S/N=-{\tt norm.ppf}(\gamma)$.
    \end{enumerate}
\item We combine the $n_c$ (e.g., three) ``Gaussian SNR'' images from each energy channel using Equation~\ref{eq:combine};
\item We look for detections in the Combined S/N map 
by searching for local maxima within the pixels satisfying the condition $SN_{tot}[SN_{tot}>-{\tt norm.ppf}(\gamma)]$.
\item We perform aperture photometry at the location of each detection, using several fixed apertures ($5$, $10$, and $15$ pixels) as well as two dynamic apertures sized at one and two times the width of the Gaussian that best fits the PSF at the detection location
\item We perform the PSF photometry using the flux estimator $\widetilde{F}=S/\mathcal{C}(\widetilde{F})$ (see Equation~\ref{eq:SF} in \citealt{Ofek+2018_PoissonNoiseMF}).
\item We use the per-observation bad pixel map provided by Chandra to flag sources whose chip coordinates (averaged over the dithering that occurs during the exposure) are near problematic pixels. Additionally, we filter the image with a delta function to reveal instrumental artifacts that may cause the appearance of ``hot'' pixels, particularly in short sub-exposures where dithering has minimal impact.
\end{enumerate}


\subsection{PSF modeling}\label{sec:psf_all}

\subsubsection{PSF calculation}\label{sec:psf}

The shape of the Chandra PSF varies significantly as a function of the position of the source in the telescope field-of-view, and as a function of the spectral energy distribution of the source. To best account for this variation, the PSF is modeled using the {\tt MARX}\footnote{https://space.mit.edu/cxc/marx/} software, a simulation tool designed to model and analyze data from the Chandra X-ray Observatory. {\tt MARX} allows users to generate realistic simulations of how X-rays propagate through Chandra's telescope, gratings, and detectors, accounting for instrumental effects and observational parameters and is particularly useful for modeling the PSF at specific detector locations and energy. We used the version of {\tt MARX} from the {\tt ciao-4.16} chandra package. 

\subsubsection{PSF library}\label{sec:library}

To improve runtime, rather than computing the PSF for each sector and observation, we can use a ``surrogate'' observation to create a library of PSF stamps on a grid of {\it detector} coordinates — a coordinate system associated with the focal plane of the Chandra instrument.

\begin{figure}
\includegraphics[width=8cm]{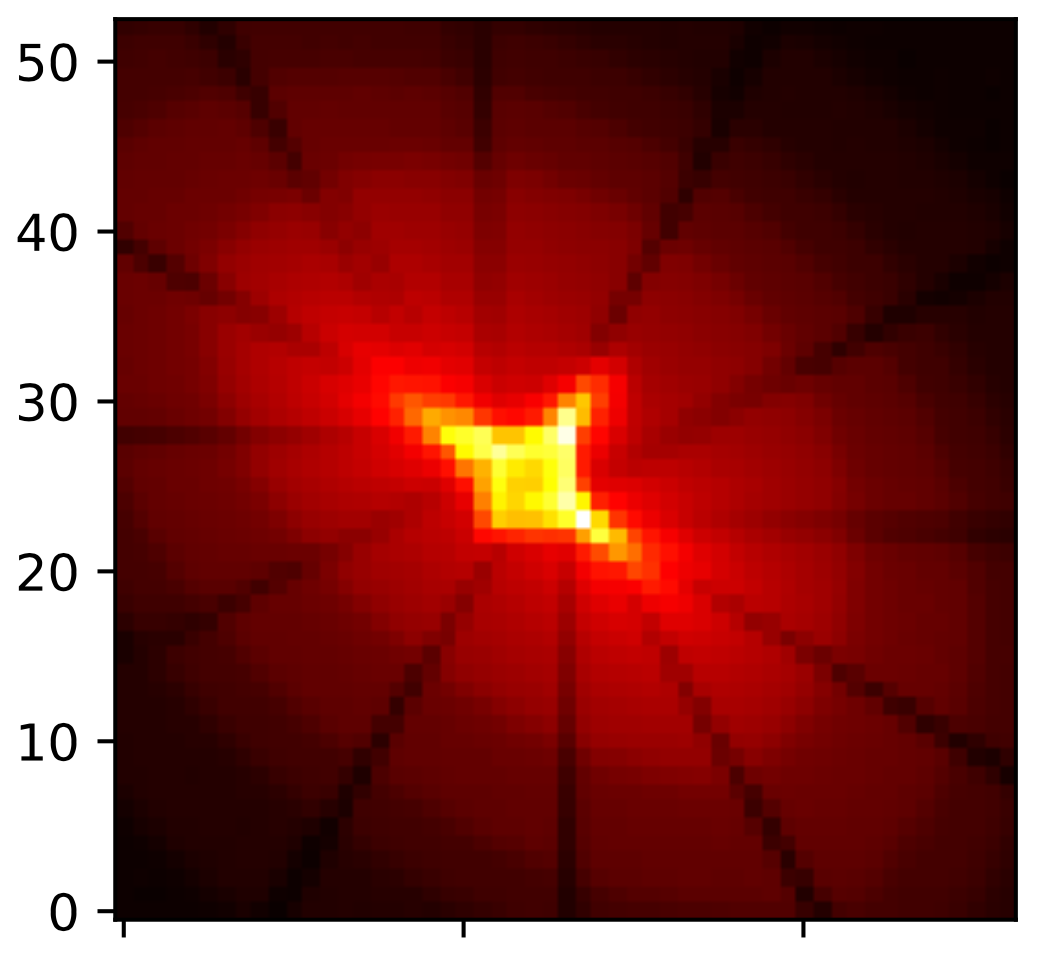}
\caption{{\bf An example of a PSF stamp modeled with {\tt MARX} for the ``surrogate'' {\tt OBSID}=$18073$}. 
Specifically, it was calculated for CCD3, $\rm{ra}=148.5963$, $\rm{dec}=69.7419$. It was centered (using a cubic-shift) and cropped to the area containing $99\%$ of the PSF, to reduce runtime.
The PSF for any given {\tt OBSID}, in the same sub-area of the detector, is deduced by rotating this stamp by the {\tt ROLL\_ANGLE} of the {\tt OBSID} minus the  {\tt ROLL\_ANGLE} of the surrogate.}
\label{fig:PSF_stamp}
\end{figure}

In the PSF stamps generated by {\tt MARX}, the PSF center does not always align with the center of the image; the RA/Dec for which the PSF was calculated may lie within a group of pixels in the central region of the image produced by {\tt MARX}. To address this, each PSF in the library is shifted to ensure it is centered on the central Chandra pixel of the sector. The PSF stamps are then cropped to include only $N\%$ of the PSF, where $N=99$ by default and is a configurable parameter of the code. This cropping step allows to further save runtime.
The appropriate PSF stamp for any random observation is then deduced by rotating the library stamp, calculated for the nearest CCD sector, by an angle $\alpha=\texttt{ROLL\_ANGLE}_{OBSID}-\texttt{ROLL\_ANGLE}_{surrogate}$. 
Figure~\ref{fig:PSF_stamp} shows an example of one of the library ``surrogate'' PSF stamps after (1) modeling the PSF with \texttt{MARX}, (2) shifting the PSF to ensure it is centered, and (3)  cropping an area that contains $99\%$ of the PSF.


The PSF library made available with the first release of {\tt X-Sifter} consists of a pre-calculated sample of $7,680$ templates ($16\times16$ sectors, in steps of 64 pixels, $\times3$ energy channels and $\times10$ CCDs; the steps size were determined by estimating the information loss due to the use of the wrong PSF).

\subsection{$S_{th}$ modeling and $\gamma-S$ relations modeling}\label{sec:modelingS}


We precompute $\mathcal{P}(S\mid H_0)$ for a range of background values and construct a library of relations between the background and $S_{th}$
for each PSF stamp in the library. The value of $S_{th}$ (step (e)) and the $\gamma-S$ relation (step (f)) for any new background value are then obtained by interpolating within this precomputed library.

\subsection{Extended sources}

To identify extended sources in images, {\tt X-swifter} provides an option to efficiently search the images by convolving the PSF with a set of extended source templates.

\section{Results and validation}\label{sec:results}
Two approaches are used to assess the performance of {\tt X-Sifter}. In \S\ref{sec:validation}, we run the pipeline on simulated images to verify that the simulated sources are successfully recovered. In \S\ref{sec:comp with CSC}, we validate the sources detected by {\tt X-Sifter} against those in the CSC2.1 catalog and compare the sensitivity of the source detection strategies.
In \S\ref{sec:FalsePos} we discuss the false positives rate.

\begin{figure*}
\centering
\includegraphics[width=8cm]{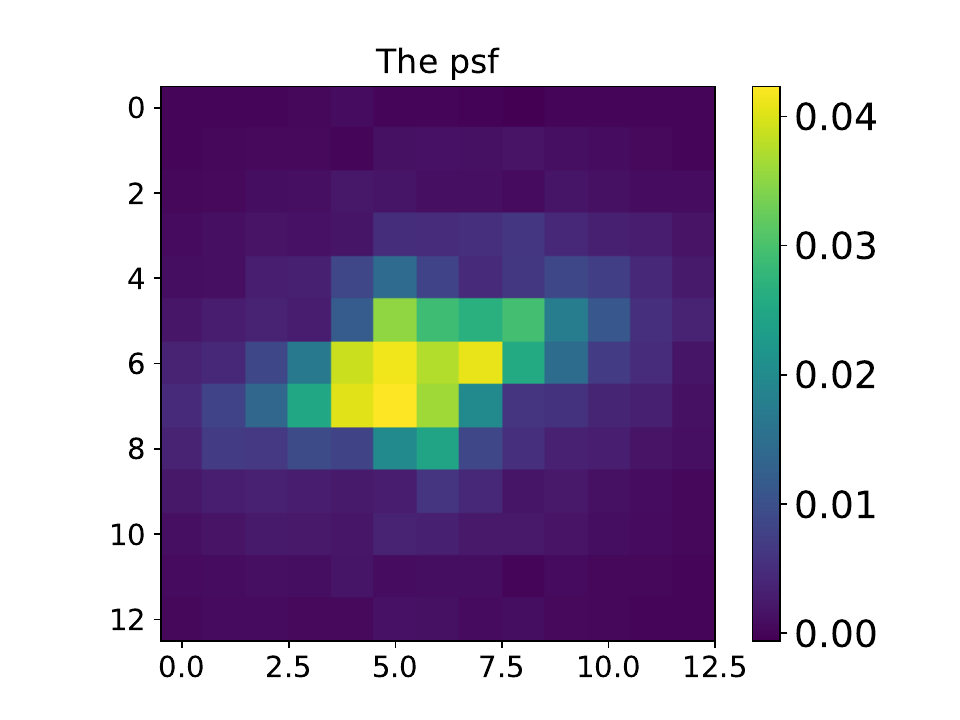}
\includegraphics[width=8cm]{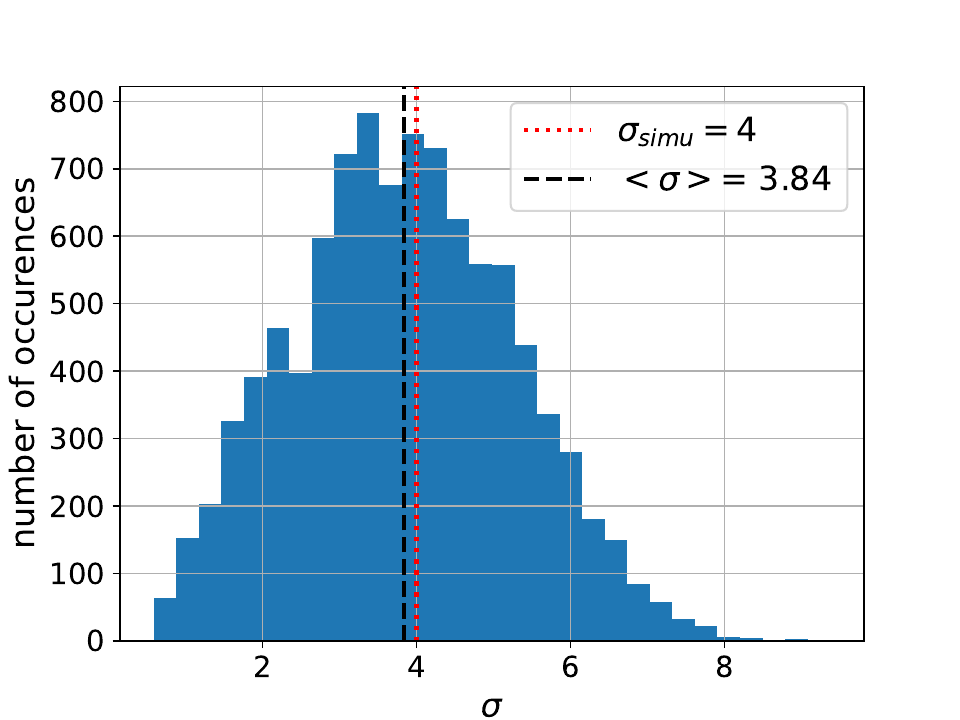}
\includegraphics[width=16cm]{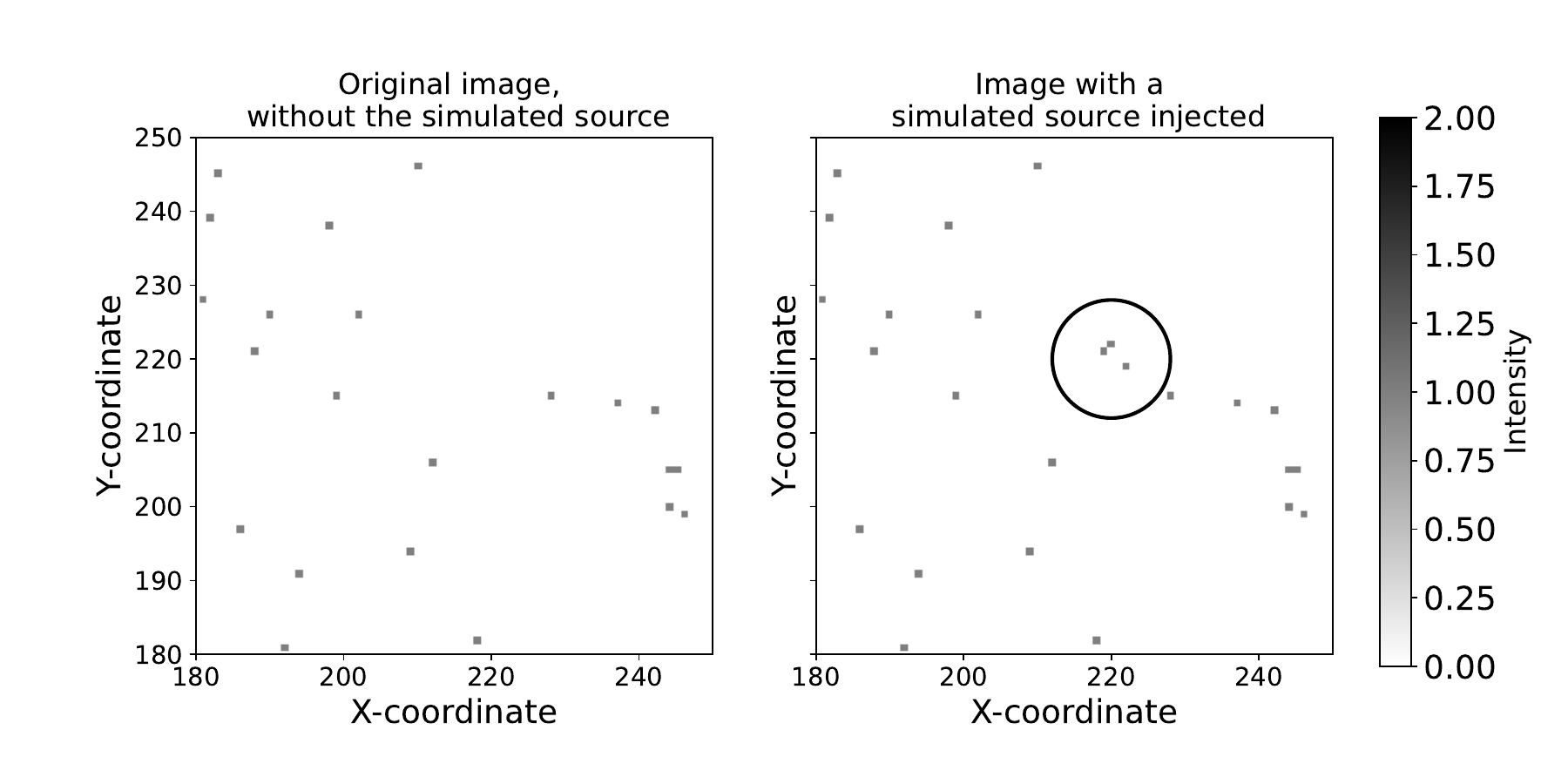}
\caption{{\bf Description of the simulations made to validate the pipeline}. For a given location on the observation \texttt{OBSID}=$18910$, the PSF was computed using {\tt MARX} (top left panel). For a chosen $\sigma=4$ level, and the background level $B$ measured for this sector, the flux of a simulated source $F_{th}$ was calculated using Equation~\ref{eq:final link} and the pixel counts $M(q-q_0)=\text{Poisson}(B + F_{th}P(q-q_0))$ for this source was injected to the existing image (the lower panel shows the image before and after the injection of the simulated source). The image was then filtered using {\tt X-Sifter} and the process repeated $10^{4}$ times, resulting in a distribution of S/N for the source detections (top right panel) peaking near $\sigma=4$, as expected.}
\label{fig:simu-semi}
\end{figure*}

\subsection{Validation of the pipeline with simulation}\label{sec:validation}
To validate the pipeline, we perform the following test, illustrated in Figure~\ref{fig:simu-semi} for \texttt{OBSID}=18910:

\begin{enumerate} \item For a given \texttt{OBSID} and a specific sector (see \S\ref{sec:algo} for a definition of a ``sector''), we measure the background level $B$ using the strategy described in \S\ref{sec:overview} (step 4(c)); \item We simulate the PSF $P$ corresponding to this sector using {\tt MARX}. The resulting PSF stamp is shown in Figure~\ref{fig:simu-semi} (top left panel); \item For a false-alarm probability $\gamma_{sim}$ corresponding to a $4\sigma$ detection, and using the value of $B$ measured in step (1) and the stamp $P$ simulated in step (2), we calculate $F_{th}$ by solving Equation~\ref{eq:final link}; \item We generate a simulated image by injecting a source modeled as $M(q-q_0) = \text{Poisson}(B + F_{th}P(q-q_0))$ at an arbitrary location $q_0$ in the real image. The simulated image is shown in Figure~\ref{fig:simu-semi} (lower panels); \item We filter the simulated image to extract the source, applying the algorithm described in \S\ref{sec:overview} (steps 4(c) to 4(j)); \item We repeat this procedure $10,000$ times and find that the distribution of signal-to-noise ($S/N$) levels peaks near $4\sigma$, consistent with the $\gamma_{sim}$ chosen (Figure~\ref{fig:simu-semi}, top right panel). \end{enumerate}

\begin{figure*}
\centering
\includegraphics[width=8cm]{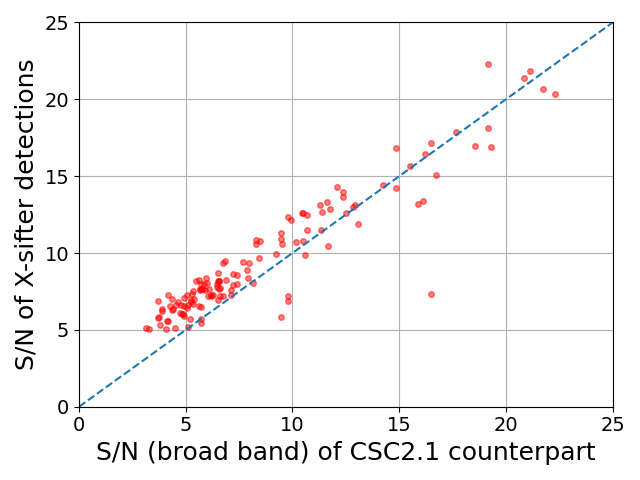}
\includegraphics[width=8cm]{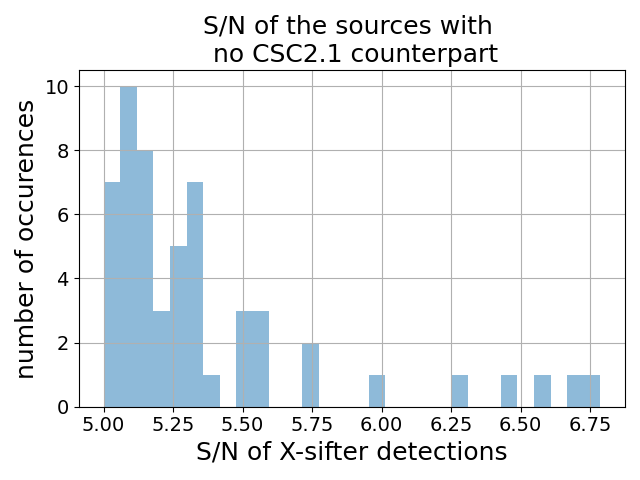}
\caption{{\bf A comparison of the source detection by \texttt{X-sifter} and by the Chandra Source Catalog CSC2.1}. Left panel: For sources detected in both catalogs, and in fields observed only once by Chandra (to ensure the sources were not detected in stacks), we show the S/N given by {\tt X-Sifter} (y-axis) and as reported in the CSC2.1 (x-axis). Right panel: the S/N measured by {\tt X-Sifter} for an additional $30\%$ of the sources which are not detected in the CSC2.1 catalog.}
\label{fig:comparaison}
\end{figure*}


\subsection{Comparison with the detection by the Chandra Source Catalog CSC2.1}\label{sec:comp with CSC}
To assess the performance of {\tt X-Sifter}, we compare the signal-to-noise (S/N) levels of detections output by {\tt X-Sifter} and by the CSC2.1 \citep{Evans2024}. The source detection strategy used in the CSC2.1 is explained in details in \cite{Evans2010} and \cite{Evans2024}. It is based on the CIAO {\tt wavedetect} algorithm \citep{Freeman2002}. In summary, {\tt wavedetect} looks for local maxima in the image data filtered (i.e. cross-correlated) with a set of Marr (“Mexican Hat”) wavelets with a range of scale sizes. The filtering is performed in separate energy channels.

We perform the comparison for a set of sources detected by both algorithms in ten observations. To ensure a fair comparison, we selected regions of the sky that were observed only once with Chandra, meaning the CSC detections were performed on a single observation rather than a stack of observations. Figure~\ref{fig:comparaison} shows that most sources detected by both algorithms were detected with a higher S/N by {\tt X-Sifter}, suggesting that {\tt X-Sifter} achieves higher sensitivity, consistent with the expectation from the Poisson Matched Filter.


The difference is particularly pronounced near the detection threshold (for $\sigma<7$), where the mean significance is increased by a factor of $\left<\rm SN_{\rm{\tt XSwifter}}/SN_{\rm CSC}\right>\approx1.3$. This corresponds to an increase in survey speed of $\left<\rm SN_{\rm{\tt XSwifter}}/SN_{\rm CSC}\right>^2\approx1.8$. As we move further from the detection threshold (i.e., for higher $\sigma$ values), the approximation used in {\tt X-Sifter}—which involves extrapolating the $\gamma-S$ exponential relation —becomes less accurate. This likely results in a misestimation of the S/N by {\tt X-Sifter}, as seen in Figure~\ref{fig:comparaison}, where its performance is less favorable compared to CSC2.1 in this regime.

Another notable point is that, for the chosen set of {\tt OBSIDs}, only two sources were detected by CSC2.1 and not by {\tt X-Sifter}. Theses two sources have a recorded S/N in CSC2.1 lower than 4.5. $55$ sources out of $179$ ($\cong 30\%$) were detected by {\tt X-sifter} and not detected by CSC2.1. Figure~\ref{fig:comparaison} shows that these sources are all very close to the detection threshold. Note that the number of detected sources is expected to increase by an order of magnitude when the temporal subdivision of the observations is activated (see \S\ref{sec:temporal}). 


\subsection{Search for false-positives}
\label{sec:FalsePos}

\begin{figure*}
\centering
\includegraphics[width=7cm]{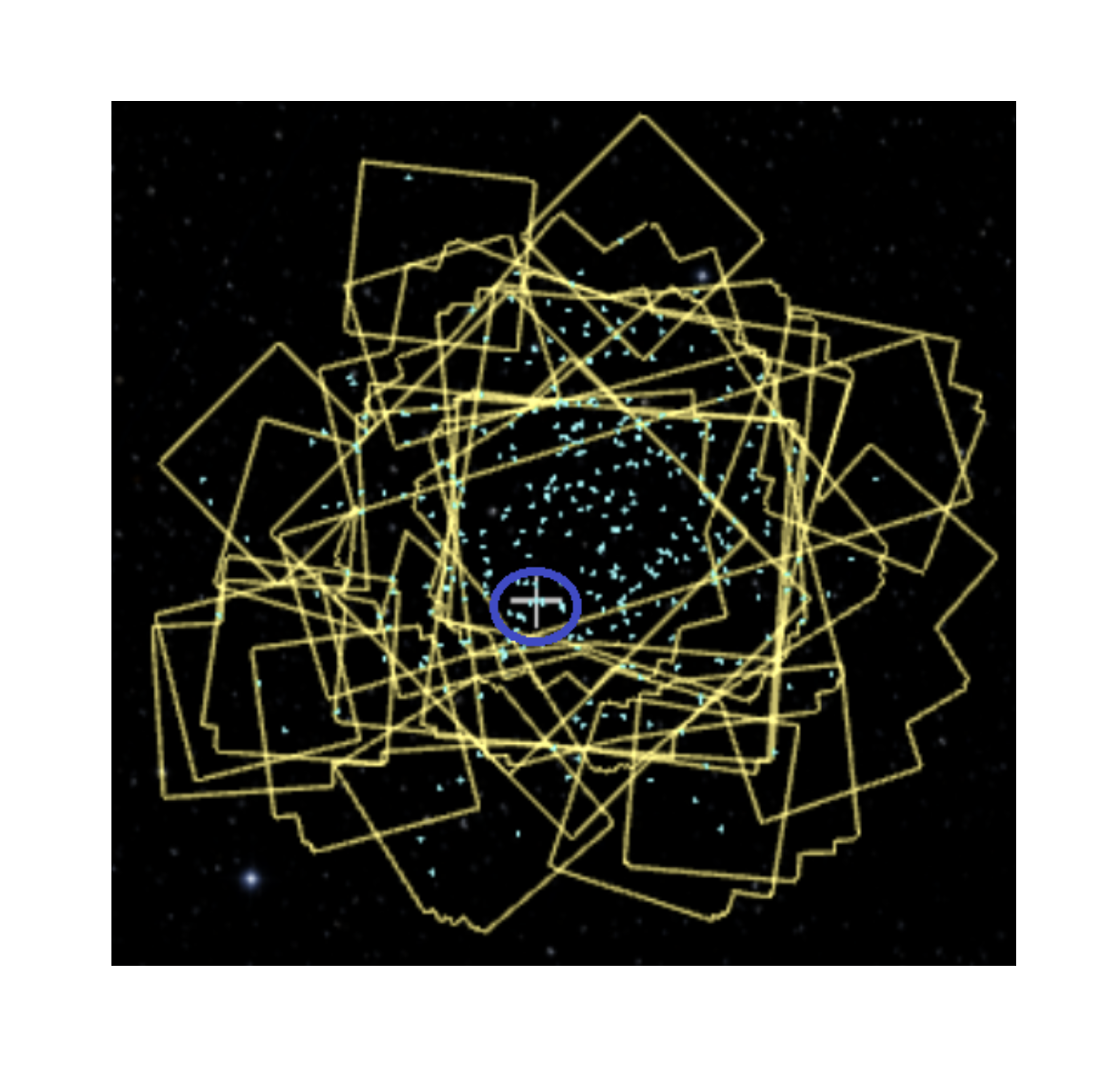}
\includegraphics[width=9cm]{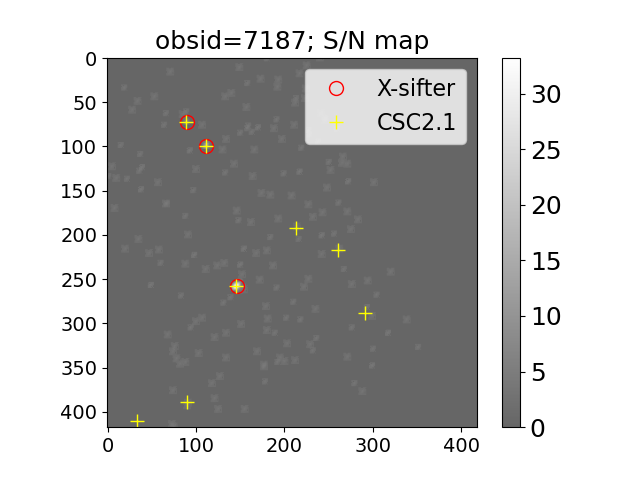}
\caption{{\bf The Search for False Positives}. Left panel: an example of a sky region covered by seven observations, including ${\tt OBSID}=21996$. The cross and circle mark the specific area used for this analysis. Right panel: a region covered by six {\tt OBSIDs}, including ${\tt OBSID}=7187$. In this specific case, as in most of the {\tt OBSIDs} included in this analysis, all sources detected by {\tt X-Sifter} are also present in the CSC2.1 catalog.
}  
\label{fig:false-positives}
\end{figure*}

To estimate the false-positive rate in the sample of {\tt X-Sifter} detections, we selected Chandra {\tt OBSIDs} from regions observed multiple times by the Chandra telescope. These regions have CSC2.1 detections performed on stacks of multiple {\tt OBSIDs}, unlike the single OBSID analysis described in \S~\ref{sec:comp with CSC}. In this scenario, we expect all sources detected by {\tt X-Sifter} to have counterparts in CSC2.1 unless the {\tt X-Sifter} algorithm produces false positives. If no false positives are present, the only other explanation for sources to be detected by {\tt X-Sifter} but not by CSC2.1 would be transient or variable events. Such transients may be buried in the background noise of the stacked data used by the CSC2.1 algorithm and only become detectable in the shorter, single exposures on which {\tt X-Sifter} operates.

We selected five {\tt OBSIDs}—$18730$, $21996$, $8470$, $7187$, and $22393$—covering coordinates observed between three and one hundred times by the Chandra telescope. Figure~\ref{fig:false-positives} illustrates one of the regions included in our analysis, along with the overlapping {\tt OBSIDs} in that area. For the selected regions, CSC2.1 identifies 48 detections, while {\tt X-Sifter} identifies 23, of which 10 lack counterparts in CSC2.1.

Upon inspecting these ten unmatched sources, we find that they are either clearly transient on the timescale of the observation (with photons concentrated within less than half of the exposure time) or undetected when running {\tt X-Sifter} on at least one of the other observations contributing to the stack, indicating that these sources varied on a timescale shorter than the duration covered by the stack. In summary, all quiescent sources detected by {\tt X-Sifter} were also identified in the CSC2.1 catalog, and the only sources detected by {\tt X=Sifter} and not in the CSC2.1 are transients. 
Given the limited test we have conducted, we have found 0 false alarms out of 23 unmatched events.
This can be used to put a 95\% confidence upper limit of 16\% on the rate of false positives. (\citealt{Gehrels1986_PoissonCI}).

\section{Conclusions}


We present the {\tt X-Sifter} pipeline, an implementation of the Poisson Matched Filter (\citealt{Ofek2018}), addressing a range of challenges associated with both the PMF formalism and the complexities of real data. In addition to the sensitivity improvement inherent to accounting for the Poisson nature of the data
— {\tt X-Sifter} incorporates several features designed to optimize detection sensitivity. These include (1) careful modeling of the PSF and background noise, accounting for their dependence on detector location and energy, and (2) temporal subdivision of the data into sub-exposures. When activated, this feature makes the pipeline more sensitive to short and elusive transients, which would otherwise become buried in the background noise of long exposures.

Tests, conducted in comparison with the CSC2.1 catalog, suggest a noticible improvement in detection capabilities. Specifically, there is a substantial ($\cong30\%$) increase in the number of sources detected near the detection limit (even before enabling the temporal subdivision feature) and a 1.3-fold increase in the significance level for sources detected with both methods, corresponding to a 1.8-fold increase in survey speed. A comparison in regions where CSC2.1 detections are based on stacked data, using a small sample of five {\tt OBSIDs}, showed that CSC2.1 includes all quiescent sources detected by {\tt X-Sifter}. This further validates our pipeline and suggest that the code does not introduce significant false positives.

A {\tt Python} implementation of the pipeline, along with several libraries designed to optimize runtime and specifically adapted for Chandra archival data, is released to the community alongside the publication of this paper through the {\tt X-Sifter} \href{https://github.com/maayane/T-Rex}{Github page}. Information on how to install and run the pipeline, as well as on its architecture, is provided on this page.
Additional tools that implement specific steps that are part of this pipeline are also available in MATLAB\footnote{\url{https://github.com/EranOfek/AstroPack}} (\citealt{Ofek2014_MAAT,Ofek+2023PASP_LAST_PipeplineI}), and they are described in a dedicated wiki page\footnote{\url{https://github.com/EranOfek/AstroPack/wiki/PhotonsList}}.


\acknowledgments

This paper is dedicated to the memory of Carmella Dan.

We thank Nicholas Lee, Kenny Glotfelty and the entire Chandra Telescope Helpdesk team for very precious advice. We thank Raffaella Marguti for useful discussion.

This research was supported by the Israeli Science Foundation (grant Nos. 2068/22 and 2751/22)

This research has made use of data obtained from the Chandra Data Archive and the Chandra Source Catalog, both provided by the Chandra X-ray Center (CXC).

This research used resources of the National Energy Research Scientific Computing Center (NERSC), a Department of Energy Office of Science User Facility using NERSC award DDR-ERCAP-0029239.

E.O.O. is grateful for support by grants from the Willner Family Leadership Institute, Madame Olga Klein-Astrachan, André Deloro Institute, Schwartz/Reisman Collaborative Science Program, Paul and Tina Gardner, The Norman E Alexander Family Foundation ULTRASAT Data Center Fund, Jonathan Beare, Israel Science Foundation, Minerva, BSF, BSF-transformative, and the Weizmann-UK.

\bibliographystyle{apj} 
\bibliography{bibliograph.bib,papers}

\end{document}